\documentclass[showpacs,amsmath,amssymb,pre,twocolumn]{revtex4}
\usepackage{graphicx}
\begin{document}
\title{Renewal,  Modulation and Superstatistics }
\author{Paolo  Allegrini$^1$}
\author{Francesco Barbi$^2$}
\author{Paolo Grigolini$^{2,3,4}$}
\author{Paolo Paradisi$^5$}
\affiliation{$^1$Istituto Nazionale di Fisica della Materia, unit\`a di Como, Via Valleggio 11, 22100 Como, Italy}
\affiliation{$^2$Dipartimento di Fisica ``E.Fermi'' - Universit\`{a} di Pisa, Largo
  Pontecorvo, 3 56127 Pisa, Italy}
\affiliation{$^3$Center for Nonlinear Science, University of North Texas, P.O. Box 311427, Denton, Texas 76203-1427, USA}
\affiliation{$^4$Istituto dei Processi Chimico Fisici del CNR, Area della Ricerca di Pisa, Via G. Moruzzi, 56124, Pisa, Italy}
\affiliation{$^5$ISAC-CNR, Sezione di Lecce, Strada Provinciale Lecce-Monteroni km 1.2 I-73100 Lecce, Italy }
\date{\today}
\begin{abstract}
  We consider two different proposals to generate a time series with the
  same non-Poisson distribution of waiting times, to which we refer to
  as renewal and modulation.  We show that, in spite of the apparent
  statistical equivalence, the two time series generate different physical
  effects. Renewal generates aging and anomalous scaling, while modulation
  yields no aging and either ordinary or anomalous diffusion, according to the
  prescription used for its generation. We argue, in fact, that the physical
  realization of modulation involves critical events, responsible for scaling.
  In conclusion, modulation rather than ruling out the action of critical
  events, sets the challenge for their identification.

\end{abstract}


\maketitle

\section{introduction}\label{introduction}

The new field of complexity is attracting the attention of an increasing
number of researchers, and it is triggering vivacious debates about its true
meaning \cite{debate1,debate2,debate3}. Here we adopt the simple minded
definition of complexity science, as the field of investigation of
multi-component systems characterized by non-Poisson statistics. On intuitive
ground, this means that we trace back the deviation from the canonical form of
equilibrium and relaxation to the breakdown of the conditions on which
Boltzmann's view is based: short-range interaction, no memory and no
cooperation. Thus, the deviation from the canonical form, which implies total
randomness, is a measure of the system complexity.

However, this definition of complexity does not touch the delicate problem of
the origin of the departure from Poisson statistics. Here we limit ourselves
to considering two different proposals, which we shall refer to as
\emph{renewal} and \emph{modulation}. We shall show that these two proposals,
although different, might lead to identical statistical results, as far as
waiting time distribution and correlation function are concerned. Thus, to a
first sight, one might be tempted to conclude that they are indistinguishable,
leaving no motivation whatsoever to prefer the one to the other. We shall
prove that it is not so, and that there exist physical properties, whose
observation allows us to distinguish the two proposals and to assess which one
is correct for the given complex process under consideration.  
We shall prove, with numerical arguments, that, whereas renewal produces aging, modulation does not. 
This is the first
result of the paper, confirming the theoretical predictions of an earlier work
\cite{grigolini}. We shall discuss also both modulation and renewal as source
of diffusion generating fluctuations. We shall argue that the modulation
scaling depends, in principle, on the special way modulation is realized. 
We shall discuss two different procedures to generate modulation and we shall
show that both produce critical events. The critical events are renewal
events, leading a wealth of secondary events, and determining their Poisson
properties. The critical events determine the scaling, which turns out to be
anomalous only if the critical events are of non-Poisson type.  This sets a
challenge for the identification of the critical events, in general.

The outline of the paper is as follows. In Section \ref{timeseries} we define the artificial time series under study in this paper, characterized by  non-Poisson time distribution. In Sections  \ref{sectionrenewal} and \ref{modulation} we  explain how to generate the non-Poisson waiting time distributions according to renewal and modulation prescriptions, respectively. In Section \ref{aging} we discuss an aging experiment that turns out to afford an efficient criterion to distinguish between modulation and renewal. Section \ref{diffusion} is devoted to studying the diffusion process generated by either renewal or modulation, with 
the surprising conclusion that the physical realization of modulation generates unexpected renewal properties. 
With Section \ref{end} we reiterate the importance of critical events, which, although ostensible in the case of  renewal theory and invisible in the case of modulation, are responsible for anomalous scaling in both cases.

\section{Time series to study}\label{timeseries}

In this paper we shall discuss numerically and theoretically the statistical
properties of a time series $\{\tau_{i}\}$ generated from within two different
approaches to complexity, \emph{renewal} and \emph{modulation}, in such a way
as to produce in both ways the same time distribution with the following form:
\begin{equation}
\label{powerlaw}
\psi(\tau) = (\mu -1) \frac{T^{\mu-1}}{(\tau+T)^{\mu}},
\end{equation}
where $\mu > 1$.  This distribution is properly normalized, and the parameter
$T$, making this normalization possible, gives information on the lapse of
time necessary to reach the time asymptotic condition when $\psi(\tau)$
becomes identical to an inverse power law. The choice of the form of Eq.
(\ref{powerlaw}) is dictated by the simplicity criterion.  This form has been
known for many years \cite{metzlernonnenmacher}, see for instance Ref.
\cite{scott}, and following Metzler and Nonnenmacher
\cite{metzlernonnenmacher} and Metzler and Klafter \cite{metzlerklafter} we
shall be referring to it as Nutting law. This form is also obtained by means
of entropy maximization from a non-extensive form of entropy \cite{tsallis}
and, for this reason, is referred to by an increasing number of researchers as
Tsallis distribution.

We shall use the sequence $\{\tau_{i}\}$ to distribute events on the time
axis. The first event occurs at $t = \tau_{1}$, the second at time $t =
\tau_{1} + \tau_{2}$, and so on. The time intervals between two consecutive
events are called \emph{laminar regions}, the reason being that the dynamical
model here under study to illustrate renewal theory is an idealization of the
celebrated Manneville map \cite{manneville}, with the time interval between
two events representing the fluid regular state.  We shall keep calling these
time intervals laminar regions even when working with modulation, even if, in
this case, as we shall see, the durations of different laminar regions are 
subtly correlated.

For the theoretical discussion of this paper, it is  useful to turn the sequence of times $\{\tau_{i}\}$ into a diffusion generating fluctuation $\xi(t)$. Each time laminar region is assigned through a coin tossing procedure either the value $W$ or the value $-W$ ($W>0$), thus creating a dichotomous fluctuation $\xi(t)$, which is interpreted as a stochastic velocity. We then shall study the time evolution of the coordinate $x(t)$, obeying the dynamic prescription 
\begin{equation}
\dot{x}=\xi.
\end{equation}
We shall try to establish to what an extent the scaling of the resulting
diffusion process depends on the origin of the fluctuation $\xi(t)$
(modulation or renewal). This is done having in mind the interesting case of
blinking quantum dots \cite{kuno}, with $W$ denoting the ``light on'' and $-W$
the ``light off'' state. It is well known in fact that the waiting time
distributions of ``light on'' and ``light off'' states depart from the Poisson
condition.

The central issue of the diffusion process under study in this paper has to do with scaling, namely, the property:

\begin{equation}
\label{scaling}
p(x,t) = \frac{1}{t^{\delta}} F\left(\frac{x}{t^{\delta}} \right),
\end{equation}
which is expected to hold true in the time asymptotic limit. The departure
from ordinary statistical mechanics is signaled by either $\delta \neq 0.5$ or
$F(y)$ departing from the Gaussian form, or by both properties. We shall find
the surprising fact that even with modulation the departure from ordinary
statistical physics is determined by crucial renewal events, which seem to be
an unavoidable consequence of any practical way we might adopt to realize
modulation.

\section{renewal} \label{sectionrenewal}
As done in earlier work \cite{-1,0,1,2,3}, we are referring ourselves to a simple  dynamic process, as a prototype of renewal model. To make this paper as self-contained as possible, let us review here this  simple model. Let us consider a particle 
moving within the interval $I = (0,1]$ driven by the following equation of
motion
\begin{equation}
\label{renewal}
\frac{d}{dt} y  = a y^{z},
\end{equation}
with
\begin{equation}
\label{inequality}
z \geq 1  ,
\end{equation}
and 
\begin{equation}
  a  > 0 .  
\end{equation}
Due to the positivity of $a$ the particle moves from the left to the
right, and any time it reaches the border $y = 1$ is injected back to a
randomly chosen position $y_{0}$, fitting the condition
\begin{equation}
0 < y_{0} < 1 .
\end{equation}
The distribution density of sojourn times, $\psi(\tau)$, is evaluated as
follows. First of all, we solve Eq. (\ref{renewal}) to determine the time
necessary for the particle to reach the border moving from a given initial
condition $y_{0}$. This time is given by
\begin{equation}
\label{renewalmadeevident}
\tau = \frac{1}{a(z-1)} \left[\frac{1}{y_{0}^{z-1}} - 1 \right].
\end{equation}
The probability for the particle to get the border in the infinitesimal
interval $[\tau, \tau + d \tau]$ is determined by
\begin{equation}
\psi(\tau) d\tau = p_{0}(y_{0})dy_{0}.
\end{equation}
We make the assumption of uniform back injection, which yields 
$p_{0}(y_{0}) = 1$.
Thus, we obtain the form of Eq. (\ref{powerlaw}) with
\begin{equation}
\mu = \frac{z}{(z-1)}
\end{equation}
and
\begin{equation}
\label{curiosityonrenewal}
T = \frac{\mu -1}{a}  .
\end{equation}
It is interesting to notice that using Eq. (\ref{powerlaw}) leads to the following 
 expression, for the mean sojourn time, $\langle \tau \rangle$

\begin{equation}\label{biggerthantwo}
\langle \tau \rangle = \frac{T}{\mu-2}.
\end{equation}
Thus, using Eq.  (\ref{curiosityonrenewal}) we express $\mu$ as follows
\begin{equation}
\label{comparetoaging}
\mu = 1 + a T =2+\frac{T}{\langle \tau \rangle}.
\end{equation}
This formula makes it evident that considering the case when the 
 invariant distribution $p_{eq}(y)$ exists, and $Ê\langle \tau \rangle< \infty$, 
 yields $\mu > 2$, which is the condition considered in this paper.

In practice, we create first the sequence $\{y_{0}(i)\}$, by means of a
succession of random drawings of numbers from within the interval $I$. Then,
using the transformation of Eq. (\ref{renewalmadeevident}), we associate
$y_{0}(i)$ with $\tau_{i}$, thereby creating the sequence $\{\tau_{i}\}$.  
We use this sequence to model, according to the renewal theory, the
physical processes under discussion in this paper. Here we limit ourselves to
using the sequence of `laminar regions' $\{\tau_{i}\}$ to distribute events on
the time axis. As earlier stated, the
first event occurs at $t = \tau_{1}$, the second at time $t = \tau_{1} +
\tau_{2}$, and so on. 


This model is inspired to the renewal theory \cite{cox}. The renewal character
of the model is made evident by Eq. (\ref{renewalmadeevident}). In fact, the
values of $y_{0}$ are randomly chosen from a uniform distribution, $0 < y_{0}
< 1$. Any drawing does not have memory of the earlier drawings. Consequently,
a laminar region does not have any memory of the earlier laminar regions. In
literature there are many examples of renewal models yielding the distribution
of Eq. (\ref{powerlaw}). Here we quote first the model illustrated by
Zaslavsky \cite{zaslavsky}.  We shall refer to this model as the hierarchical
islands trap. It has to do with the condition of weak Hamiltonian chaos, when
regular islands are surrounded by a chaotic sea. The surface of separation is
characterized by the island-around-island phenomenon: any iteration of a
zooming process, yielding an increasing magnification, makes the same
structures appear at smaller and smaller space and time scales. In other
words, a particle that reaches this separation surface through diffusion in
the chaotic sea undergoes a stochastic motion that can be adequately described
by a master equation corresponding to an unbounded chain of
states, $|w\rangle_{i}$, with $i = 1, 2, \cdots, \infty$.  The first state of
this chain, $i = 1$, is a doorway state, establishing a connection between the
chaotic sea and the power law generating chain. The particle in this state can
either jump back to the chaotic sea or forward to the second state of the
chain, $i=2$. From the second state the particle can either jump back to the
doorway state, $i = 1$, or forward to the third state, $i = 3$. As the
particle moves forward, namely $i$ increases, the jump rate becomes smaller
and smaller. Given the fact that both motion forward, increasing values of
$i$, and backward, decreasing values of $i$, are possible, the particle,
sooner or later will jump back to the chaotic sea. A renormalization group
approach \cite{zaslavsky} yields a distribution of waiting times
corresponding, in the time asymptotic limit, to an inverse power law, and thus
to a theoretical prediction compatible with the choice of Eq.
(\ref{powerlaw}). The renewal nature of the process is ensured by the fact
that, once the particle jumped back to the chaotic sea, the memory of the long
sojourn in the power law generating chain is lost.

In the specific case of blinking quantum dots, a model of the same kind as
that of Ref. \cite{zaslavsky} has been proposed by the authors of Ref.
\cite{legitimate}. This model, in turn, is formally equivalent to that
proposed years ago by Bouchaud \cite{bouchaud1,bouchaud2}, to explain the
dynamics of glassy systems. These two models are significant examples of a
wider category of renewal models, including the hierarchical islands trap
model illustrated by Zaslavsky \cite{zaslavsky}.  It is straightforward to
adapt the picture of hierarchical islands trap to the physics of blinking
quantum dots. In this case, we say that the electron makes a jump from a state
where spontaneous emission of light is possible, to a state that can be
thought of as the doorway state of a region where light emission is quenched.
From this state the electron can jump either back or forward, the forward
direction corresponding to a deeper and deeper embedding within the ``light
off'' region. The forward and backward jumping rates become smaller and smaller
with an increased embedding. The electron can come back to the doorway state
and from there to the ``light on''region.  We can adopt an analogous model for
the ``ligth off'' state. In the real case of blinking quantum dots this is
a non-Poisson distribution, different from the non-Poisson distribution of the
times of sojourn in the ``light off'' state. In this paper, for simplicity, we
make the assumption that the two distribution are identical.
This generates a sequel of times of alternate sojourn in the ``light on'' and
``light off'' state, with no correlation whatsoever among themselves. We shall
see that the modulation approach apparently yields the same result, but it
generates a subtle form of correlation among different sojourn times.

\section{modulation theory}\label{modulation}
We define as modulation theory any approach to non-Poisson distribution based
on the modulation of Poisson processes. For instance, a double-well potential
under the influence of white noise yields the Poisson distribution of the time
of sojourn in the two wells \cite{earlier}. In the case of a symmetric
double-well potential we have
\begin{equation}
\label{kramers}
\psi(t) = \lambda \exp(-\lambda t).
\end{equation}
The parameter $\lambda$ is determined by the Arrhenius formula
\begin{equation}
\lambda = k \exp\left(- \frac{Q}{k_{B} T}\right).
\end{equation}
In the case when either the barrier intensity $Q$ \cite{earlier} or
temperature $T$ \cite{fonseca} are slowly modulated, the resulting waiting
time distribution becomes a superposition of infinitely many exponentials. At
least since the important work of Shlesinger and Hughes \cite{important}, and
probably earlier, it is known that a superposition of infinitely many
exponentially decaying functions can generate an inverse power law. This, by
itself, does not qualify the theory as modulation. It depends on the criterion
adopted to generate the sequence $\{\tau_{i}\}$, mentioned in Section \ref{timeseries}. If values of $\tau_{i}$, with different labels $i$ are selected
from different exponential distribution, the resulting process is no doubt
renewal. To make it become a form of modulation theory, we have to draw a
large sequence of time values, with the index $i$ moving from $i_{\lambda}$ to $i = i_{\lambda} + N_{d}$, with $N_{d} >> 1$, from the same Poisson distribution $\psi(\tau) = \lambda \exp(-\lambda \tau)$.

In recent times, the term superstatistics has been coined \cite {cohen} to
denote an approach to non-Poisson statistics, of any form, not only the
Nutting (Tsallis) form, as in the original work of Beck \cite{beck}. We note
that Cohen points out explicitly \cite{cohen} that the time scale to change
from a Poisson distribution to another must be much larger than the time
scale of each Poisson process. Thus, we can qualify superstatistics as a form
of modulation.  Therefore, from now on we shall indifferently refer to this
approach to complexity either as modulation or superstatistics.

In conclusion, according to the modulation theory we write the waiting time
distribution $\psi(t)$ under the following form
\begin{equation}
\label{ideal}
\psi(\tau) = \int d\lambda \Pi(\lambda) \lambda \exp(-\lambda t),
\end{equation}
where $\Pi(\lambda)$ is the $\Gamma$ distribution of order $\mu-1$ 
given by
\begin{equation}
\Pi(\lambda)=\frac{T^{\mu-1}}{\Gamma(\mu-1)}\lambda^{\mu-2}\exp{(-\lambda T)}.
\label{paradiso}
\end{equation}
This formula was proposed by Beck \cite{beck} and used in a later work
\cite{bologna}.

We make the assumption
of being able to generate time series with no computer time and computer
memory limitation. Of course, this is an ideal condition, and in practice we
shall have to deal with the numerical limits of the mathematical recipe that
we adopt here to understand modulation. The reader might imagine that we have
a box with infinitely many labeled balls. The label of any ball is a given
number $\lambda$. There are many balls with the same $\lambda$, so as to fit
the probability density of Eq. (\ref{paradiso}). We randomly draw the balls
from the box and after reading the label we place the ball back in the box. Of
course, this procedure implies that we are working with discrete rather than
continuous numbers. However, we make the assumption that it is possible to
freely increase the ball number so as to come arbitrarily close to the
continuous probability density of Eq. (\ref{paradiso}).

After creating the sequence $\{\lambda_{j}\}$, we create the sequence
$\{\tau_{i}\}$ with the following protocol. For any number $\lambda_{j}$, the
reader must imagine that we have available a box with another set of
infinitely many balls. Each ball is labeled with a number $\tau$, and in this
case the distribution density is given by $\psi(\tau) = \lambda \exp(- \lambda
\tau)$.

To realize modulation, we adopt two different prescriptions:

{\bf Prescription N.1} We create a sequence $\{\tau^{(j)}_{i}\}$ by making
$N_{d}$ drawing from this box. Notice that according to the arguments of Cohen
\cite{cohen} for modulation to be identified with superstatistics, it is
necessary to make $N_{d}$ very large, virtually infinite. Notice that the
correlation function of the fluctuation $\xi$, 
\begin{equation}
\Phi_{\xi}(\tau) \equiv \frac {1}{W^2} \langle \xi(t) \xi(t+\tau) \rangle,
\end{equation} 
for any fixed
$\lambda$ is equal to $\exp(-\lambda \tau)$. We note also that the smaller
$\lambda$ the larger is the time interval corresponding to it. Consequently,
for a proper definition of the effect of modulation on $\Phi_{\xi}(\tau)$, we
have to use the statistical weight $\Pi(\lambda)/\lambda$ \cite{bologna},
which yields
\begin{equation}
\label{equalequal}
\Phi_{\xi}(\tau) = \frac{\int d\lambda \frac{\Pi(\lambda)}{\lambda} \exp(-\lambda \tau)}{\int d\lambda \frac{\Pi(\lambda)}{\lambda} }.
\end{equation}
This means that the waiting time distribution $\psi(\tau)$, of eq.
(\ref{ideal}) is proportional to the second time derivative of Eq.
(\ref{equalequal}), this being a known consequence of renewal theory
\cite{geisel}.  Thus, for $\mu > 2$, modulation and renewal not only yield the
same $\psi(\tau)$, but also the same correlation function. 

As a consequence, studying the second moment of the diffusing variable $x$,
through the average over the distribution $p(x,t)$ of the diffusing process is
not sufficient to discriminate between renewal and modulation with a fixed
number of drawings.  We shall see, however, that the aging experiment of
Section \ref{aging} allows us to distinguish between modulation and renewal,
in spite of the fact that, quite surprisingly, the adoption of this modulation
prescription generates in the time asymptotic limit the same anomalous scaling
as the renewal process. This is so because this kind of prescription generates
by its own non-Poisson crucial events. These crucial events are imbedded in a
sea of Poisson events that make them essentially invisible: the non-Poisson
aging is annihilated by an overwhelming quantity of Poisson events that, yet,
are not robust enough as to make the asymptotic scaling become ordinary. This
delicate aspect is discussed in detail in Section \ref{diffusion}.

{\bf Prescription N.2}. In this case we keep drawing the waiting time numbers
from the same Poisson distribution for a fixed amount of time $T_{d}$. In this
specific case
\begin{equation}
\label{equaletime}
\Phi_{\xi}(\tau) = \int d\lambda \Pi(\lambda)  \exp(-\lambda \tau).
\end{equation}
In this case the correlation function is obtained from the first time
derivative of $\psi(t)$. Thus, in this case when $\mu > 2$ the correlation
function is integrable. The scaling of the diffusion process is expected 
to be that of standard diffusion, namely $\delta = 0.5$, while 
$F(y)$ of Eq. (\ref{scaling}) is a Gaussian function.

\section{Aging effects in renewal and modulation theories}\label{aging}

Aging of renewal processes is a term indicating the non-stationary nature 
of the probability distribution in the complete phase space 
${x, \xi, {\bf y}}$, 
where the vector ${\bf y}$ stands for the variables responsible
for the dynamics of $\xi$, and hence responsible for $\psi(t)$.
When starting from an off-equilibrium condition, the probability 
distribution of waiting times is affected by the
evolution of the phase space distribution. As a consequence, 
the distribution of waiting times for the first event changes with the
time $t_a$, which is defined as the time delay between \emph{preparation}
and \emph{observation}.
Let us explain this concept in some detail. We assume that
preparation sets the system in a brand new, or ``young''
condition. This means  a Gibbs ensemble with  all the trajectories located 
at the beginning of a laminar region.
Let us define $t=0$ as the time at which the observation of this system starts.
The first measured waiting time is denoted by 
$\tau_1$. The first waiting time, at variance with the observation of the successive
waiting times, does not necessarily correspond to the total time duration of a laminar region. In fact, the
first laminar region could have started at some unknown time
$t=-t_a$, so that the first real waiting time is, let us say, given by 
$\tau_1+t_a$. 
It should be noted that this is the key property generating aging and memory in the non-Poisson case. In fact, in the Poisson case the truncation of the laminar regions stemming from an observation process with delay respect to preparation
has the effect of altering both short and long-time laminar regions in the same way. As a consequence, a renewal process with Poisson statistics does not show
aging. In the non-Poisson case, on the contrary, the short-time laminar region suffer from this truncation process more than the long-time laminar regions. Thus the distribution intensity at short times is lowered and at large times is enhanced, so generating a slower decay. The experimental observation allows us to establish at which time the system was prepared, a form of long-standing memory that yet is compatible with the fact that the laminar regions showing up in the future do not have any memory of those that occurred in the past.  In the model of Section \ref{sectionrenewal}, this process of memory erasure is caused  by the
random back injection. 

In this section, we illustrate, with the help of a numerical example, how
aging effects are deeply related to the renewal character of the process,
whereas aging is annihilated by the slow modulation. Monte Carlo
simulations of a renewal process with inverse-power-law distribution of
waiting times, Eq. (\ref{powerlaw}), are performed using the approach proposed
in Section \ref{sectionrenewal}, {\it i.e.}, we generate a sequence
$\{\tau_{i}\}$ by using Eq. (\ref{renewalmadeevident}). 

As far as modulation theory is concerned, the numerical treatment is a little
bit more delicate.  The first step in the Monte Carlo simulation consists in
generating a sequence of random numbers $\lambda$ in agreement with the
$\Gamma$ distribution $\Pi(\lambda)$ given in Eq. (\ref{paradiso}).  An
approach similar to the one used in Section \ref{sectionrenewal} cannot be
applied, because there is no simple analytical expression for the cumulative
function related to $\Gamma$ distributions.  As a consequence, it is not
possible to obtain a simple explicit expression for the function relating the
random number $\lambda$ to a uniform random number in the interval $I =
[0,1]$. For this reason, we decided to use the rejection method suggested by
Ref. \cite{numericalrecipes}. This method is based on the use of a majorant
function $f(\lambda)$, {\it i.e.}, a function that, for each point, takes
values slightly greater than the corresponding values of the $\Gamma$
distribution. The general idea is to draw random points in the plane,
uniformly distributed under the graph of $f(\lambda)$. Then, the points under
the graph of the $\Gamma$ distribution are accepted and the others rejected.
The random points are drawn as follows.  First of all, we draw a random number
$\lambda$ distributed as the probability distribution $f(\lambda)/A$, where $A
> 1$ is the area under the graph of $f(\lambda)$. This is easily done if
$f(\lambda)$ is chosen with a simple analytical expression, in order to apply
the method of Section \ref{sectionrenewal}. Then, given $\lambda$, a uniform
random number $\zeta$ is drawn in the interval $[0,f(\lambda)]$.  The
coordinates $(\lambda,\zeta)$ define the random point. Finally, if $\zeta$ is
smaller than the corresponding value $\Pi(\lambda)$, then $\lambda$ is
accepted, otherwise it is rejected.  Without loss of generality we fix $T=1$
and we chose the following majorant function:
\begin{equation}\label{rejection}
f(\lambda)=\left\{ \begin{array}{ll}
        \gamma e^{-\lambda}  & \mbox{if $\lambda \leq 1$};\\
        \frac{\eta}{\lambda^2}  & \mbox{if $\lambda>1$}.\end{array} \right.
\end{equation}
According to Ref. \cite{numericalrecipes}, in order to obtain a certain number of accepted numbers, rejection
method requires a greater amount of random drawings, depending on the choice
of the majorant function. We found that, given the majorant function described
by Eq. (\ref{rejection}), with $\eta=1.7$ and $\gamma=\eta\cdot e$
($e$ is the Neper's number), the total number of drawings is about the
double of the accepted drawings. Furthermore, we compared the histograms
computed from the sequence of simulated random numbers with the relative
probability distribution $\Pi(\lambda)$ given in Eq. (\ref{paradiso}), with
excellent agreement.
Following Prescription N. 1, for each number $\lambda$ we have to draw $N_d$
numbers from the exponential distribution described in Eq. (\ref{kramers}),
representing a sequence of waiting times.
This is easily obtained with the same standard method described in Section
\ref{sectionrenewal}.
Now, we used this approach to generate trajectories, i.e., artificial
sequences of random waiting times. We compared the results for the aging
analysis of trajectories characterized by the same exponent $\mu$ of the
power-law, but generated from the renewal process and from modulation
processes with different $N_d$. 

\begin{figure}
  \includegraphics[angle = 270, scale = 0.3]{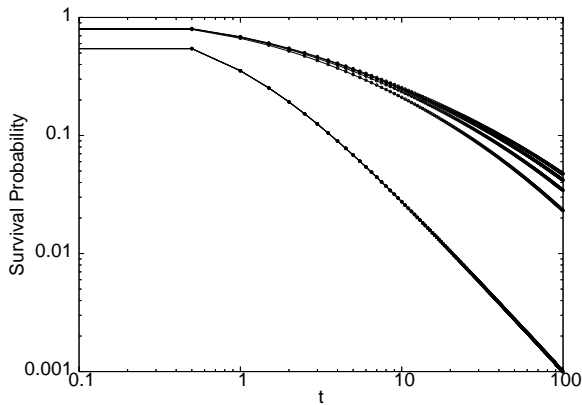}
  \caption{Comparison between the function $\Psi_{t_{a}}$ of  a renewal
   process (continuous lines) and the $\Psi_{t_{a}}$ of a modulation process with
   $N_d=1$ (dots) at different values of $t_a = 0, 50, 100, 150, 200$
   (from the lower to the upper curves).}  
  \label{ren-nd1}
\end{figure}

\begin{figure}
  \includegraphics[angle = 270, scale = 0.3]{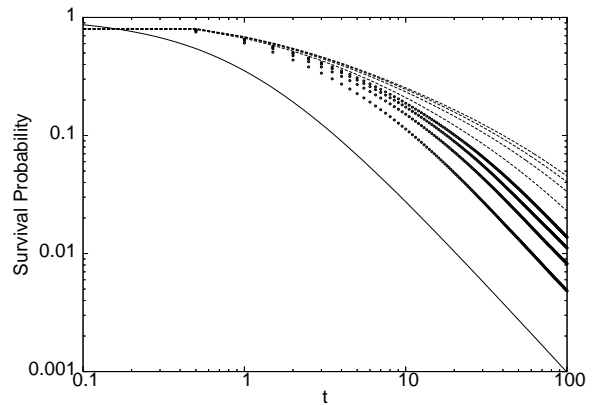}
  \caption{Comparison between the function $\Psi_{t_{a}}$
   of  a renewal process (dashed lines) and the function $\Psi_{t_{a}}$ of a modulation process with
   $N_d=10$ (dots),  at different values of $t_a = 50, 100, 150, 200$
   (from the lower to the upper curves). The lowest curve represents the
   brand new function $\Psi(t)$.}
  \label{aging_ren-nd10}
\end{figure}

\begin{figure}
  \includegraphics[angle = 270, scale = 0.3]{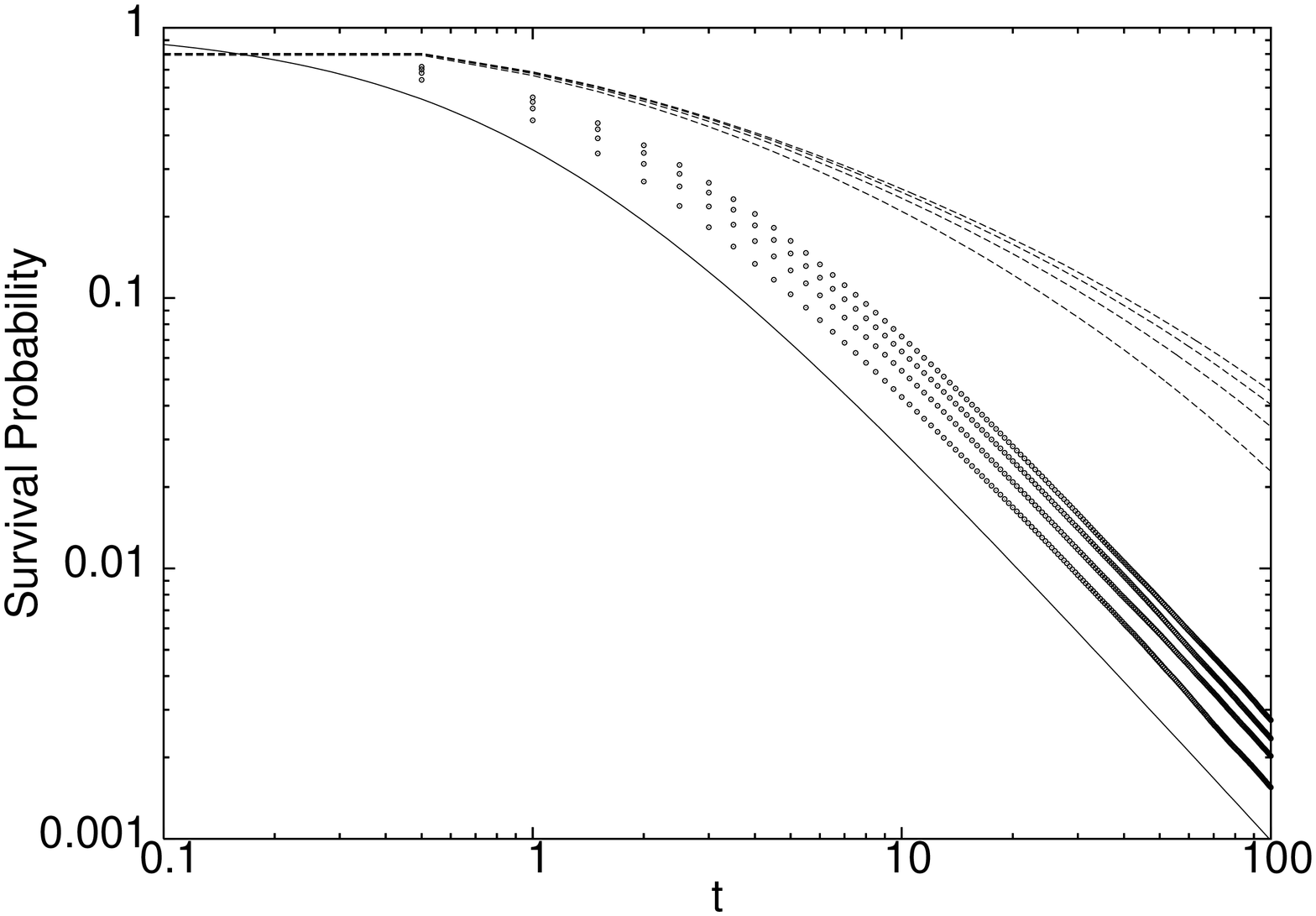}
  \caption{The same as Fig. \ref{aging_ren-nd10}, but with $N_d=100$.}  
  \label{aging_ren-nd100}
\end{figure}

\begin{figure}[!h]
  \includegraphics[angle = 270, scale = 0.3]{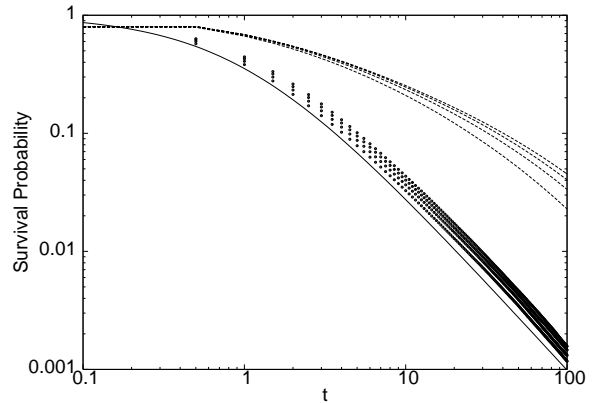}
  \caption{The same as Fig. \ref{aging_ren-nd10}, but with $N_d=500$.}  
  \label{aging_ren-nd500}
\end{figure}

We perform the aging analysis in the
following way. Given the sequence of waiting times and an aging time $t_a$,
we compute the truncated waiting times, {\it i.e.}, the difference between
each waiting time and $t_a$. When the waiting time is shorter than $t_a$,
then we take the successive waiting times until their sum exceeds $t_a$.
Then the truncated waiting time is defined as the difference between this sum
and $t_a$. In this way, we obtain a sequence of truncated waiting times
characterized by some {\it aged} probability distribution $\psi_{t_a} (\tau)$.
Aging is revealed if this distribution changes with $t_a$.
Note that the modulation process resting on Prescription N.1,  for $N_d=1$ is expected to
coincide with the prediction of renewal theory. To confirm this important property by means of numerical 
calculation, in Fig. \ref{ren-nd1} we make a comparison between the renewal  and the
modulation process,  with $N_d=1$. Rather than measuring the function $\psi_{t_{a}}$, we evaluate the 
{\it aged} survival probability $\Psi_{t_{a}}$, which is related to $\psi_{t_{a}}$ by the following relation:
\begin{equation}
\Psi_{t_a}(\tau)=\int_\tau^\infty \psi_{t_a} (\tau') d\tau'=
1-\int_0^\tau \psi_{t_a} (\tau') d\tau'.
\label{survival}
\end{equation}
As expected, the aged survival probabilities with the same $t_a$, relative to
the renewal and the modulation process respectively, coincide. In particular
this happens for the brand new survival probabilities $\Psi(t) \equiv
\Psi_{t_{a} = 0} $.  The Survival Probability Functions $\Psi_{t_a}(\tau)$ all
decrease with $\tau$ and do not cross.  Furthermore, for each fixed $\tau$,
$\Psi_{t_a}(\tau)$ are increasing with $t_a$.
The comparison of Figs. \ref{aging_ren-nd10}, \ref{aging_ren-nd100} and
\ref{aging_ren-nd500} reveals the {\it lack of aging} in the modulation theory
with respect to renewal theory, which yields the maximum aging effect
for the smallest value $N_d=1$, coinciding with the renewal condition. In the limiting case $N_d \rightarrow
\infty$ the aging effect is annihilated.
%
%




\section{diffusion}\label{diffusion}

We address the problem of diffusion generated by modulation using two distinct
procedures: prescriptions N.1 and N. 2 of Section {\ref{sectionrenewal}. We
  shall use the continuous-time random walk to derive the asymptotic scaling
  of the process. The problem of diffusion with renewal has been the object of earlier
  work \cite{jacopo} to which we refer the interested reader to appreciate the
  difference between renewal and modulation on this specific issue.  Here we
  limit ourselves to noticing that the {\em L\'{e}vy scaling}, namely $\delta
  = 1/(\mu-1)$ and $F$ of Eq. (\ref{scaling}) being a symmetric L\'evy
  $(\mu-1)$-stable function, is a property of the central part of the spreading
  distribution, which is limited by ballistic peaks forcing it to become
  multi-scaling.

\subsection{Continuous Time Random Walk: fixed number of drawings}
We now address the issue of revealing the crucial events corresponding to the
use of prescription N. 1. At the same time we plan to give a more accurate
definition of crucial event, a property that so far has not yet clearly
defined. The procedure that we plan to apply is as follows.  We study the
diffusion process from $t=0$ to a given time $t>0$, by assuming that for the
particle to reach the position $x$ at time $t$ it is necessary to have $n$
crucial events. By a crucial event we mean the drawing of a given value of
$\lambda$ from the random distribution $\Pi(\lambda)$.  Let us explain why the
time of this drawing signals the occurrence of a crucial event. The concept of
event implies unpredictability, a property shared by the random drawing of a
given $\lambda$ and by the successive $N_{d}$ random drawings of the times
$\tau_{i}$'s from
\begin{equation}\label{psilambda}
\psi_{\lambda} (\tau) = \lambda \exp(- \lambda  \tau )
\end{equation}
However, the property of being \emph{crucial} is assigned only to the 
 drawing of a given $\lambda$. In fact, all the successive $N_{d}$ time 
 drawings adopt the same Poisson prescription, dictated by this 
 leading drawing. It is remarkable that the sequence $\{\tau_{i}\}$ 
 would pass a random test, based on the ordinary correlation function, 
 due to the random drawing. However, as we have seen in Section V, 
 this sequence does not pass the aging test, which is a proof of the 
 subtle correlation created by the persistent use of the same Poisson 
 distribution for a large time span.
 
 We define as $P(x,t)$ the probability of moving by a quantity $x$, either
 positive or negative, during time $t$, as an effect of the $N_{d}$ drawings
 from the same Poisson distribution. Analogously, with the symbol
 $P^{(n)}(x,t)$ we denote the probability that the particle moves by the
 quantity $x$, as a result of $n$ crucial events, occurring in such a way as
 to ensure that the particle moves by the quantity $x$ exactly in a time $t$.
 Note that after the occurrence of these $n$ crucial events, the particle is
 in general located at a distance $x'$ from the origin different from $x$.
 With $n+1$ events the particle might overshoot that position. Thus the
 particle might make the remaining trip using the value of $\lambda$ drawn at
 the last crucial event, occurred at a time $t' < t$. In other words, we
 calculate $p(x,t)$, {\em i.e.\/}, the probability density of finding the
 particle in position $x$ at time $t$ using the formula
\begin{eqnarray}
&&p(x,t) = \nonumber \\
&&=\sum_{n=1}^{\infty} \int_{0}^{\infty} dt' \int_{-\infty}^{+\infty}
P^{(n)}(x',t') N(x-x',t-t') \nonumber \\ 
&&+ N(x,t). 
\end{eqnarray}
Here the quantity $N(x,t)$ denotes the
probability of moving by a quantity $x$ in a time $t$ with no crucial event
occurring. The Fourier-Laplace transform of $p(x,t)$, $\hat p(k,u)$, is given
by the expression
\begin{equation}
\label{towardsscaling}
\hat p(k,u) = \frac{1}{1 - \hat P(k,u)} \hat N(k,u).
\end{equation}
This expression is of crucial importance to determine the resulting scaling.
To reach this important result we have to define first the function
$\psi^{(i)}(x,t)$.  This is the probability of moving by the quantity $x$ in time
$t$, with $i$ time drawings from the same Poisson distribution. This important
function reads
\begin{equation}
\label{importantprescription}
\psi^{(i)}(x,t) = \int_{0}^{\infty} d\lambda \psi_{\lambda}^{(i)}(x,t) \Pi(\lambda),
\end{equation}
where 
$\psi_{\lambda}^{(i)}(x,t)$ denotes the probability of moving by $x$ in time
$t$ as a result of drawing $i$ numbers from a {\em given} Poisson distribution, thereby corresponding to the same value of the parameter $\lambda$. 
The functions $\psi_{\lambda}^{(i)}(x,t)$ obey the following recursion relation: 
\begin{equation}
\label{generic}
\psi_{\lambda}^{(i)}(x,t) = \int_{0}^{t} dt' \int_{-\infty}^{+\infty} dx' \psi_{\lambda}^{(i-1)}(x',t') \psi_{\lambda}^{(1)}(x-x',t-t'),
\end{equation}
with
\begin{equation}
\label{onedrawing}
\psi_{\lambda}^{(1)}(x,t) \equiv \lambda \exp(-\lambda t) \frac{1}{2} \left[\delta(x-Wt) + \delta(x+ Wt)\right].
\end{equation}
In conclusion, the physical meaning of the function $\psi^{(i)}(x,t)$ of
(\ref{importantprescription}) corresponds to adopting the following procedure. With probability $\Pi(\lambda)$ we
draw a given $\lambda$. Then, we draw $i$ values of time intervals, and we move
the particle by a quantity $x$ with these $i$ drawings.

With these prescriptions we express the quantity $P(x,t) \equiv P^{(1)}(x,t)$
as follows
\begin{equation}
P^{(1)}(x,t) = \psi^{(N_{d})}(x,t). 
\end{equation}
In other words, the transition by the quantity $x$ in time $t$ generated by a
single crucial event is derived from Eq. (\ref{generic}) by setting $i =
N_{d}$. Let us remind the reader that our approach to modulation is based on
drawing $N_{d}$ times from the same Poisson distribution.

Finally, we must express $N(x,t)$. As earlier mentioned, the function $N(x,t)$
denotes the probability of moving the particle by a quantity $x$ in time $t$
with no crucial event involved. Thus,
\begin{equation}
N(x,t) = \int_{0}^{+\infty} d\lambda  \Pi(\lambda) N_{\lambda}(x,t),
\end{equation}
and
\begin{eqnarray}\label{nocrucial}
&&N_{\lambda}(x,t) = \\
&&=\sum_{i=1}^{N_{d}} \int_{0}^{t} dt' \int_{-\infty}^{+\infty} dx' \psi_{\lambda}^{(i-1)}(x',t') \Psi_{\lambda} (x-x',t-t'),\nonumber
\end{eqnarray}
where with $\psi_{\lambda}^{(0)}(x,t)$ we denote the initial condition
$\delta(x)\delta(t)$
In
(\ref{nocrucial}) there are no crucial events involved because the maximum
numbers of time drawings is $N_{d}-1$, not enough to generate a new crucial
event, if we start at time $t=0$ with a crucial event. As a result of this
partial number of drawings the particle reaches position $x'$ in time $t'$.
The remainder portion of space $x-x'$ in the remainder portion of time $t-t'$
is traveled with no further time drawing, according to the prescription
\begin{equation}
\Psi_{\lambda}(x,t)  = \frac{\left[\delta(x+Wt) + \delta(x-Wt)\right]}{2} 
\int_{t}^{+\infty} dt' e^{-\lambda t'}.
\end{equation}

To establish the scaling produced by the modulation approach, we have to evaluate the Fourier-Laplace transform of $\psi^{(i)}(x,t)$, $\Psi_{\lambda}(x,t)$ and $P(x,t)$. Using the convolution theorem we have:
\begin{equation}
\label{almostcritical}
\hat \psi^{(i)}(k,u) = \int_{0}^{+\infty} d\lambda\Pi(\lambda) \left[\frac{\lambda(u+\lambda)}{(u+\lambda)^{2} + k^{2} W^{2}}\right]^{i}
\end{equation}
and
\begin{equation}
\hat \Psi_{\lambda}(k,u) = \frac{(u+\lambda)}{(u+\lambda)^{2} + k^{2} W^{2}}.
\end{equation}
From Eq. (\ref{almostcritical}) we derive immediately
\begin{equation}
\label{crucial}
\hat P(k,u) = \int_{0}^{+\infty} d\lambda\Pi(\lambda) \left[\frac{\lambda(u+\lambda)}{(u+\lambda)^{2} + k^{2} W^{2}}\right]^{N_{d}}.
\end{equation}

The evaluation of the asymptotic properties, $u \rightarrow 0$ and $k
\rightarrow 0$, is essential to assess the system scaling. If we make the
assumption of sending $u$ and $k$ to $0$ first, and then integrating over
$\lambda$, we get
\begin{equation}
\label{asGibbs}
\hat p (k,u) = \frac{1}{u + k^{2} W^{2} \left\langle\frac{1}{\lambda}\right\rangle_{\lambda}},
\end{equation}
where the angle brackets with the subscript $\lambda$ indicate an average over the weight
$\Pi(\lambda)/\lambda$ as in Eq. (\ref{equalequal}). Eq. (\ref{asGibbs}),
yielding
\begin{equation}
\label{prediction1}
\delta = 0.5,
\end{equation}
is equivalent to considering an ensemble of Poisson processes, each of them
characterized by a fixed value of $\lambda$, selected from the distribution
$\Pi(\lambda)/\lambda$ and then kept fixed forever.
This suggests that the resulting scaling obeys the ordinary 
 prescription $\delta=0.5$. This is a correct property if $\mu > 3$. 
 We note, however, that when $\mu < 3$, with $\mu >2$, due to the 
 constraint of Eq. (\ref{biggerthantwo}),
 \begin{equation}
\left\langle \frac{1}{\lambda} \right\rangle_{\lambda} \propto \int_0^{\infty} d\lambda \frac{\Pi(\lambda)}{\lambda^2}=+\infty,
\end{equation}
casting some doubts on the conclusion that $\delta = 0.5$.


This is an incorrect conclusion indeed.  In fact, with some algebra we find that
\begin{eqnarray}
\label{secondterm}
\psi_{\lambda}^{(2)}(x,t) = 
\frac{\lambda^{2} t \exp(-\lambda t) }{4} \hspace{1cm}  \\
\times \left\{ \left[\delta(x-Wt) + \delta(x+Wt)\right] + \frac{\theta(Wt-|x|)}{Wt} \right\},\nonumber
\end{eqnarray}
where $\theta$ denotes the step Heaviside function,
and
\begin{eqnarray}
\label{thirdterm}
\psi_{\lambda}^{(3)}(x,t) = \frac{\lambda^{3}t^2\exp(-\lambda t) }{16}  \hspace{1cm}\\
\times \left\{\left[\delta(x-Wt) + \delta(x+Wt)\right] +3 \frac{\theta(Wt-|x|)}{Wt} \right\}.\nonumber
\end{eqnarray}
The reader should compare these higher order distributions with the expression
of Eq. (\ref{onedrawing}). We see that increasing the number of drawing has
the effect of weakening the intensity of the ballistic component that, in the
case of only one drawing, is known to generate anomalous diffusion
\cite{zumofenklafter}.  In fact, for only one drawing, using Eq.
(\ref{onedrawing}), we have
\begin{equation}
\psi^{(1)}(t) = \frac{1}{2} \left[\delta(x-Wt) + \delta(x+Wt)\right] \psi_{\mu}(t),
\end{equation}
where
\begin{equation}
\psi_{\mu}(t) \equiv (\mu-1) \frac{T^{(\mu-1)}}{(t+T)^{\mu}}.
\end{equation}
Thus, with only one drawing, we obtain the same result as that provided by the
renewal theory. Notice that, with only one drawing, the dynamic process stems
from the motion of two ballistic peaks (the two delta of Dirac of Eq. (28)).
We see that already with $N_d=2$, a contribution embedded between the to
ballistic terms appear. We make a first estimation of scaling, based on
neglecting these non-ballistic contributions. It is therefore evident that
this prediction is not exact.

To predict the effect of increasing $N_{d}$ we notice that for $t \rightarrow
\infty$ the function $\psi_{\lambda}^{(i)}(x,t)$ will tend to a Brownian
motion with a diffusion coefficient $\propto{1/\lambda}$. This process, namely
a renewal process with a single value of $\lambda$ is known to be described by
the celebrated telegrapher's equation, whose solution is due to Cattaneo
\cite{cattaneo}.  If we define $\Lambda_{\lambda}^{(i)}(t)$ the intensity of
the ballistic peaks of $\psi_{\lambda}^{(i)}(x,t)$, namely within a single
value of $\lambda$, and $\Lambda^{(i)}(t)$ the intensity of the peaks for the
averaged distribution $\psi^{i}(x,t)$, we find for the former the asymptotic
expression
\begin{equation}
\Lambda_{\lambda}^{(i)}(x,t) = \frac{1}{2^{i}}\left[\delta(x-Wt) + \delta(x+Wt)\right]\frac{\lambda^{i}t^{i-1}}{i-1!} \exp(-\lambda t).
\end{equation}
In this expression we identify three factors with a specific meaning. The
factor $2^{-i}$ is a signature of the fact that for each ballistic
contribution we tossed the coin $i$ times, always getting the same result. The
second term $[\delta(x-Wt) + \delta(x+Wt)]$ means that we are considering only
the ballistic motion.  Finally in the third term,
$\lambda^{i}t^{i-1}\exp(-\lambda t)/(i-1)!$ one can easily recognize the
Poisson distribution of obtaining the $i$-th success at the $t$-th trial, in a
urn model. This indicates that the elapsed time is generated by $i$ distinct
drawings from the same Poisson distribution, with the same $\lambda$.  Let us
average $\Lambda_{\lambda}^{(i)}(x,t)$ over the weight $\Pi(\lambda)$
and write 
\begin{eqnarray}
\Lambda^{(i)}(x,t) = \frac{1}{(i-1)!2^i} \left[\delta(x-Wt) +
  \delta(x+Wt)\right] \nonumber \\ 
\times \left[\frac{\partial^{i-1}}
{\partial\epsilon^{i-1}}\int_{0}^{\infty}d\lambda \Pi(\lambda)\lambda \exp(\lambda \epsilon t)\right]_{\epsilon = -1}.
\end{eqnarray}
From this form we obtain
\begin{eqnarray}
\Lambda^{(i)}(x,t) = \left[\delta(x-Wt) + \delta(x+Wt)\right]\\
\times \frac{(\mu-1) \mu \cdots (\mu + i -2) T^{\mu-1}
  t^{i-1}}{(i-1)!2^i(t+T)^{\mu+i-1}}. \nonumber
\end{eqnarray}
The next step rests on replacing $i$ with $N_{d}$, thereby defining the
ballistic contribution to $P(x,t)$, which is, as earlier pointed out, the
probability of travelling by the quantity $x$, in the positive, $x> 0$, or
negative, $x< 0$, direction, in time $t$. At this stage, we are ready to find
the waiting time distribution of the times of sojourn between two consecutive
critical events, denoted by as as $\psi_{critical}(t)$. 
We obtain this waiting time distribution, concerning crucial events, 
by integrating $\Lambda^{(N_{d})}(x,t)$ over $x$, a procedure that yields 
\begin{eqnarray}
\psi_{critical}(t) &=& \frac{1}{(N_d -1)! 2^{N_d-1}} 
(\mu-1)\mu \cdots (\mu + N_{d} -2) \nonumber \\ 
&&\times \frac{T^{\mu-1} t^{N_{d}-1}}{(t+T)^{\mu+N_d-1}}.
\end{eqnarray}
We see that the distribution of times of sojourn between two critical events reaches a maximum at
\begin{equation}
t = \frac{(N_{d}-1)T }{\mu}.
\end{equation}
Thus, increasing $N_{d}$ has the effect of postponing the transition to the
asymptotic regime. However, when the asymptotic regime is eventually reached,
we see the emergence of an inverse power law with index $\mu$.  Due to the
renewal character of the critical events, we get as a scaling $\delta$, the
following expression
\begin{equation}
\label{anomalousscaling}
\delta = \frac{1}{\mu-1},
\end{equation}
which corresponds  to a L\'evy diffusion process \cite{jacopo}.

As earlier remarked, this prediction is not exact, given the fact that we have neglected the non-ballistic contributions to diffusion. We have considered only  the delta of Dirac contributions, whose
weight moreover tends to decrease upon increasing $N_{d}$. We wonder which might be the
physical effect of the non-ballistic contributions that we have neglected. 

We estimate the role of the neglected terms by means of a heuristic approach,
based on approximating the telegrapher's diffusion with the ordinary Brownian
diffusion. We denote this Brownian process with $p_\lambda(x,t)$, expressing
the probability of traveling by the quantity $x$ in a time $t$, under the
condition that a given $\lambda$ has been drawn at the beginning of the
laminar region, at time $t=0$, and no further drawing of $\lambda$ has
occurred.  We mentioned already that the diffusion coefficient is inversely
proportional to $\lambda$. It is easy to find that
\begin{equation}\label{brown}
 p_\lambda(x,t) = \frac{1}{(4\pi W^2t /\lambda)^{1/2}} \cdot \exp\left(\frac{-x^2\cdot \lambda}{4W^2 t}\right).
\end{equation}
To find the distribution $M(x)$ for the space traveled in a time $t_c$ between
two events, we assume $t=t_c\approx N_d/\lambda$ and we average over $\lambda$,
namely, we write
\begin{eqnarray}\label{emofics}
 &&M(x) = \frac{T^{\mu-1}}{\Gamma(\mu-1)} \times \\
&&\int_{0}^{\infty}\frac{\lambda}{(4\pi W^2N_d)^{1/2}}
\exp\left(\frac{-x^2 \lambda^2}{4W^2 N_d}\right) \lambda^{\mu-2} \exp\left(-\lambda T\right) 
  d\lambda \nonumber.
\end{eqnarray}
Since the first moment $\langle \tau_{critical} \rangle$ of $\psi_{critical}$
is finite, following \cite{jacopo} we assume that asymptotically, namely
after a very large number of critical events, the central part of both
$P(x,t)$ and $p(x,t)$ is approximated by a jumping process, with 
jumps  of 
intensity $M(x)$, occurring at regular times, namely
the time interval between two 
 consecutive jumps, is fixed to be equal to $\langle \tau_{critical} \rangle$.

It is well known that if the second moment of $M(x)$ is finite, the 
 system falls in the basin of attraction of Gauss diffusion and 
 $\delta = 0.5$. If, on the contrary, the second moment of $x$ is 
 divergent, the system falls in the basin of attraction of L\'{e}vy 
 diffusion, the diffusion process being alpha stable. This means that 
 to recover finite moments we have to study the fractional moments 
 $\langle |x|^{\alpha}\rangle$, with $\alpha < 2$. Note that the index $\mu$ of the 
 L\'{e}vy inverse power law is $\mu = \alpha + 1$. Thus, a heuristic way 
 to establish if the process is of L\'evy kind or not, rests on the 
 evaluation of the fractional moment $\langle |x|^{\alpha}\rangle$, with the weight 
 $M(x)$ of Eq. (\ref{emofics}). Then, for any $\alpha$ we must establish which is the
 threshold value of $\mu$. By threshold value of $\mu$ we mean the 
 value of $\mu$, below which the fractional moment $\langle |x|^{\alpha}\rangle$ diverges.

 In other words, we set $\alpha < 2$, and we look for the 
 minimum Êvalue of $\mu$ ensuring
 \begin{equation}\label{alphamoment}
 \int_{-\infty}^{+\infty} dx |x|^{\alpha} 
\int_{0}^{+\infty}d \lambda \lambda^{\mu-1} 
\exp\left[ \frac{-(x \lambda)^2}{4W^2N_d}  -\lambda T \right] 
< \infty.
\end{equation}
We reverse the integration order and use the fact that the Gaussian 
moments are always finite, to get
\begin{equation}\label{gaussianmoment}
 \int_{-\infty}^{+\infty} dx |x|^{\alpha} \lambda^{\alpha+1}
\exp\left[ \frac{-(x \lambda)^2}{4W^2N_d} \right] = C.
\end{equation}
Note that $C$ is independent of of $\lambda$. 
Thus the condition of (\ref{gaussianmoment}) becomes
\begin{equation}
\int_{0}^{+\infty}d \lambda \lambda^{\mu-\alpha-2} 
\exp\left( -\lambda T \right) < \infty.
\end{equation}
Note that the only source of divergence for this integral is the 
 limiting condition $\lambda \rightarrow 0$. In fact the exponential 
 factor $\exp(-\lambda T)$ ensures convergence for all values of 
 $\lambda$. To ensure the convergence of this integral we must focus 
 on the short values of $\lambda$. We see that the integral 
 convergence is ensured by $\mu > \alpha + 1$. According to the earlier 
 illustrated criterion, this yields for the scaling $\delta = 
 1/\alpha$, $\delta = 1/(\mu-1)$, thereby confirming the L\'{e}vy 
 scaling of Eq. (\ref{anomalousscaling}).
 
 \subsection{Exact evaluation of asymptotic properties}
The L\'{e}vy scaling emerging from a sea of Poisson events, which 
 annihilate, as we have seen in Section V, any sign of renewal aging, 
 is a disconcerting property. The arguments based on the analysis of 
 $M(x)$ afford already a compelling proof of this surprising fact. 
 However, as a form of double check, we plan to reach here the same 
 conclusion through the study of $\hat p(k,0)$, which, according to the 
 L\'{e}vy theory \cite{MontrollandBruce},
 should be $p(k,0) \approx 1 - const \cdot k^{\alpha}$. Using (25) we get 
 this asymptotic property provided that we show that $\hat P(k,0) 
 \approx 1 - const \cdot k^{\alpha}$. Note that we have to prove that $\alpha 
 = \mu -1$.
 For this purpose we focus on Eq. (\ref{crucial}), we set $u = 0$, and we obtain:
\begin{equation}\label{crucialinzero}
 \hat P(k,0) = \frac{k^{(\mu-1)}}{\Gamma(\mu-1)}\int_{0}^{\infty} y^{\mu-2} e^{-Tky}\biggl(\frac{y^2}{y^2+1}\biggr)^N dy.
\end{equation}
At this point, we use the binomial expansion and the useful property:
\begin{equation}
\biggl(\frac{y^2}{y^2+1}\biggr)^N 
= 
\frac
     {y^{2N}}
     { \sum_{i=0}^N 
\binom{N}{i}
 y^{2i}} 
= 
1-\frac{\sum_{i=0}^{N-1} 
\binom{N}{i}
 y^{2i}}
       {\sum_{i=0}^N 
\binom{N}{i}
 y^{2i}}.
\end{equation}
Let us plug this expression into the integral, and divide it
in two parts:
\begin{eqnarray}
\frac{k^{(\mu-1)}}{\Gamma(\mu-1)} &\times& \\
\biggl[\int_{0}^{\infty}
y^{\mu-2} e^{-Tky} dy &-&
\int_{0}^{\infty}\frac{\sum_{i=0}^{N-1} \binom{N}{i}
y^{2i}}{\sum_{i=0}^N \binom{N}{i} 
 y^{2i}}
y^{\mu-2} e^{-Tky} dy\biggr].\nonumber
\end{eqnarray}
The first term gives $\Gamma(\mu-1)/k^{(\mu-1)}$, and therefore the
unity term in the zero-th order of $\hat P(k,0)$.
Looking at the second term, note that the integrand does not diverge in zero, at the infinity either: the integral can be considered as a Laplace
transform (in $Tk$) of a finite quantity. So, at this first
order, we obtain a constant.

In conclusion,  for the integral (\ref{crucialinzero}) 
we obtain the expression
$1-const \cdot k^{\mu-1}$, which is the same first order expansion
as that stemming from the renewal process, thus yielding (\ref{anomalousscaling}). 
We think that at this stage the 
 validity of Eq. (\ref{anomalousscaling}) is widely proved. In spite of the approximation 
 made to get the L\'{e}vy scaling, Eq. (\ref{anomalousscaling}), this property is correct 
 and is expected to show up in the time asymptotic limit.


\subsection{Continuous time random walk: fixed time $T_{d}$ for the action of the same $\lambda$}
We have focused so far our attention on the modulation prescription N. 
 1. Let us consider now the modulation prescription N. 2. 
 In the case where we keep drawings the times $\tau$ from the same $\lambda$
for a time $T_{d}$, assumed to be very large, we can apply the following
argument. The probability for the particle to travel by the quantity $x$ is
given by
\begin{equation}\label{j}
M(x) \equiv p(x,T_{d}) = \int_{0}^{+\infty} d\lambda \Pi(\lambda) p_{\lambda}(x,T_{d}),
\end{equation}
where $p(x,T_{d})$ denotes the probability of travelling by the 
 quantity $x$, in the positive, $x>0$, or negative, $x< 0$, direction, for a 
 time $T_{d}$, throughout which the random times $\tau_{i}$ has been 
 always drawn from the same Poisson distribution of Eq. (\ref{psilambda}), with a 
 fixed value of $\lambda$.
 Note that also with prescription N. 2 the crucial events correspond to 
 the time when a new value of $\lambda$ is drawn. As a consequence of 
 the random choice of $\lambda$, Êas well as of the random choice of 
 $\tau_{i}$, the function $M(x)$ must be interpreted as a distribution 
 of totally uncorrelated numbers. Therefore, as done earlier, we have 
 to focus now our attention on the fractional moments of 
 $M(x)$. Notice that, with prescription N. 2, the jumps occur at regular 
 times, separated by the fixed distance $T_{d}$, a property that makes 
 easier and even more rigorous our scaling evaluation.
 
For this prescription we have 
\begin{equation}
 p_\lambda(x,T_d) = \frac{1}{(4\pi W^2T_d /\lambda)^{1/2}}  
\exp\left(\frac{-x^2 \lambda}{4W^2 T_d}\right).
\end{equation}
Substituting (\ref{brown}) into equation (\ref{j}), we obtain:
\begin{equation}
 p(x,T_d) = \frac{\int_{0}^{\infty} \lambda^{1/2} \exp\left(\frac{-x^2 \lambda}{4W^2 T_d}-\lambda T\right)\lambda^{\mu-2}T^{\mu-1} d\lambda}{(4\pi W^2T_d)^{1/2}\Gamma(\mu-1)} 
 .
 \end{equation}
Let us consider for simplicity $T=1$:
 \begin{eqnarray}
p(x,T_d) = \frac{1}{\Gamma(\mu-1)}  
\frac{1}{(4\pi W^2T_d)^{1/2}}  \nonumber \\ 
\times  \int_{0}^{\infty}
\exp \left[-\left(\frac{x^2}{4W^2 T_d} + 1\right) \lambda\right] 
 \lambda^{\mu-3/2}  d\lambda.
 \end{eqnarray}
 Now, using integral $3.381(4)$ of Ref.\cite{keyreference}, 
\begin{equation}
\int_{0}^{\infty} x^{\nu-1} e^{-\mu x} dx = \frac{1}{\mu^{\nu}} \Gamma (\nu),
\end{equation}
with $\Re \{\mu\} > 0$ and $\Re \{\nu\} > 0$, we obtain
\begin{equation}
M(x) = \frac{\Gamma(\mu-1/2)}{\Gamma(\mu-1)}  \frac{1}{(4\pi
 W^2T_d)^{1/2}} \bigg(\frac{4W^2T_d}{x^2 +
 4W^2T_d}\bigg)^{\mu-1/2}
\end{equation}

We see that for $\mu > 2$, the second moment of $M$ is finite. Thus, 
in this case the standard central limit theorem applies, and consequently
modulation yields
\begin{equation}
\delta = 0.5
\end{equation}
rather than
\begin{equation}
\delta = \frac{1}{\mu -1},
\end{equation}
the scaling predicted by the renewal condition. 
Notice that in this case the correlation function $\Phi_{\xi}(t)$ is given by
\begin{equation}
\label{corelatiofunction}
\Phi_{\xi}(t) = \left(\frac{T}{T + t}\right)^{\mu-1}.
\end{equation}
This means that the correlation function is integrable. In Section VI D we
shall illustrate two examples of diffusion equations based on the correlation
function $\Phi_{\xi}(t)$ rather than on the critical events, visible or
invisible. The adoption of the correlation function $\Phi_{\xi}(t)$ is an
important step of the ordinary procedures of statistical mechanics, which
ignore the concept of critical event, and the concept of event as well. The
emphasis of the conventional approaches is based in fact on the density time
evolution, and the stochastic trajectories, with abrupt jumps is only
indirectly taken into account. We shall see that modulation agrees with these
theories with no critical events, only when diffusion becomes ordinary. Even
in this case, however, resting only on the experimental evaluation of $\psi(t)$
would be misleading, since it would create the impression that non-ordinary
statistical conditions apply, while the asymptotic behavior of the system is
ordinary.

\subsection{Equations for processes with no crucial events}\label{kubo}
We have seen that the physical realization of modulation implies the occurrence of critical events, these critical events being determined by the procedure we adopt to change the parameter $\lambda$ with time. 
As an ideal example of equations without critical events we
have in mind the following two generalized diffusion equations
\begin{equation}
\label{proposal1}
\frac{\partial}{\partial t} p(x,t) = W^{2}\int_{0}^{t} dt^{\prime} \Phi_{\xi}(t^{\prime}) \frac{\partial^{2}}{\partial x^{2}} p(x,t-t^{\prime})
\end{equation}
and 
\begin{equation}
\label{proposal2}
\frac{\partial}{\partial t} p(x,t) = W^{2}\int_{0}^{t} dt^{\prime} \Phi_{\xi}(t^{\prime}) \frac{\partial^{2}}{\partial x^{2}} p(x,t)
\end{equation}
Both equations are generated in such a way as to fulfill the dynamical
constraint
\begin{equation}
\label{scalingfromhere}
\frac{d}{dt} \langle x^{2}(t) \rangle = D(t),
\end{equation}
where
\begin{equation}
D(t) = W^{2} \int_{0}^{t} dt^{\prime} \Phi_{\xi}(t^{\prime}).
\end{equation}
The latter equation is proven to yield a Gaussian form for $p(x,t)$, whose
width is given by $\langle x^{2}(t) \rangle$ of Eq. (\ref{scalingfromhere}),
thereby yielding the scaling

\begin{equation}
\label{scalingwithnoevent}
\delta =  1 - \frac{\beta}{2}.
\end{equation}
Note that this scaling is determined by the correlation function
$\Phi_{\xi}(t)$, which has the form
\begin{equation}
\Phi_{\xi}(t) = \left[\frac{T}{t+ T}\right]^{\beta},
\end{equation}
with $\beta = \mu-2$ or $\mu -1$, according to whether prescription N. 1 or
N. 2 is used.  

It is interesting to notice that also the former equation, Eq.
(\ref{proposal1}) yields the scaling of Eq. (\ref{scalingwithnoevent}), as
proved by the detailed analysis of Ref. \cite{bolognone}.  It is convenient to
notice that these two equations have a different origin, sharing only the
assumption that no critical event exists. Both equations refer to the simple
diffusion process described by $dx/dt = \xi(t)$. Eq. (\ref{proposal2}) is
exact under the assumption that $\xi(t)$ is a Gaussian stochastic variable.
For its derivation, see, for instance, Ref. \cite{annunz} and references therein.
Eq.  (\ref{proposal1}) refers to the case where the variable $\xi$ is a
dichotomous process, and it is derived by using a Liouville-like approach
ignoring the existence of trajectories with sporadic but abrupt jumps \cite{bolognone}.
It is interesting to notice that both equations produce the same anomalous
scaling of Eq. (\ref{scalingwithnoevent}) even when the correlation function is not
integrable. We have seen that the prescription N.  1, generates anomalous
scaling, of the form of L\'evy, thus departing from the scaling of these
theories, with the same correlation function, but with no critical events.
With prescription N. 2 the critical events fall in the Gauss basin of
attraction, the correlation function is integrable, and modulation produces
the same scaling as the theories without critical events, but in this case the
diffusion process is ordinary. This suggest that the emergence of anomalous
scaling, of L\'evy kind, is always a signature of critical events, either
visible or invisible.




\section{concluding remarks} \label{end}
Let us discuss which are the main results emerging from this paper.  First of
all, we found that the waiting time distribution is not, by itself, an
unambiguous indicator of complexity. This confirms the conclusion of earlier
work \cite{memorybeyondmemory} from within a new perspective made attractive by the modulation, or
superstatistics, approach to complexity \cite{cohen,beck}. It is remarkable that the
time sequence generated according to the modulation prescription would pass
the randomness test, based on the use of the correlation function method. In
fact, as we have seen, the correlation enforced by modulation refers to the
persistent use of the same Poisson prescription to generate waiting times that
are otherwise totally uncorrelated. We have seen that the adoption of the
aging experiment of Section V bypasses the limits of the conventional methods
of analysis, since it establishes beyond any doubt the difference between the
renewal and the modulation character of a sequence of times. This is a fact of
some physical importance for the physics of blinking quantum dots \cite{kuno}. The
work of Brockmann \emph{et al.} \cite{brokmann} has already established aging in the
intermittent fluorescence of these new materials, and the more recent work of
Ref. \cite{bianco} confirms that this aging is due to the renewal character of
the process. Thus, we conclude, in agreement with Refs.
\cite{grigolini,bianco}, that the dynamic process responsible for intermittent
fluorescence must be built up on the basis of a renewal perspective.

Of course, the fact that superstatistics is ruled out as a proper perspective
for the physics of blinking quantum dots does not mean, in any way, that there
are no interesting complex processes in nature that might lay their roots on
this attractive perspective. This remark is related to the second important
result of this paper. Modulation, behind superstatitics, corresponds to a slow
physical process whose realization does not rest on a unique procedure.
However, it is a plausible that any realization of modulation might generate
its own critical events.  These crucial events are embedded in a Poisson sea,
exerting a camouflage action, which makes it extremely difficult, to reveal
their existence. A few years ago the authors of Ref. \cite{giacomo}
established a technique of analysis of time series, through scaling
evaluation, which was proved by later work \cite{memorybeyondmemory} to be an efficient method to
reveal the statistical properties of critical events that, in spite of being
invisible, namely, not even being recorded, determine the properties of the
visible events, the events that are recorded. This paper proves that the
physical realization of modulation rests on another class of invisible events,
which determine the stochastic rather than the deterministic properties of the
visible events.  This sets the challenge for the detection of this new kind of
invisible events. In Section VI we have seen that these invisible events
establish the anomalous scaling of the process. On the basis of the earlier
work of Ref. \cite{jacopo} we think that this anomalous scaling refers to the central
part of the probability distribution, and that the resulting anomalous
diffusion is multiscaling. There are good reasons to believe that with
suitable developments along the lines of the work of Refs. \cite{giacomo,jacopo}, it
will be possible to address successfully this challenging problem. This is
left as a subject of future work.

Finally, we want to remark that this paper establishes another important
result, concerning the deep connection between critical events and anomalous
scaling. There are theories that apparently produce complexity without
involving critical events. In Section \ref{kubo} we have illustrated two diffusion
equations generating anomalous scaling without involving critical events.
Superstatistics is not compatible with any of them, but in the trivial case of
ordinary diffusion. The reason is made clear by the results of this paper: the
practical realization of superstatistics implies the occurrence of critical
events, which might be either Poissonian or non Poissonian. In the second case
a correspondence is generated between scaling and waiting time distribution,
but the anomalous scaling is determined by these non-Poissonian critical
events, rather than by the waiting time distribution. In fact, as proved in
Section VI, the scaling is not determined by $\psi(t)$, the visible waiting
time distribution: the anomalous scaling is determined by
$\psi_{critical}(t)$, namely the distribution of the time distances between an
invisible critical event and the next.  This makes the role of critical events
even more important, and it makes the renewal perspective emerge also from
within superstatitics, a theoretical perspective that seems to be apparently
alternative to renewal theory.

 \section*{Acknowledgments}
 PG thankfully acknowledges the Welch foundation for financial support 
 through grant \# 70525.

\end{document}